\documentclass{article}
\usepackage{spconf,amsmath,graphicx}
\usepackage{amssymb}
\usepackage{multirow}

\usepackage{caption}
\usepackage{subcaption}
\usepackage{flushend}

\title{Frequency-Temporal Attention Network for Singing Melody Extraction}
\name{Shuai Yu $^1$, Xiaoheng Sun $^1$, Yi Yu $^2$ and Wei Li* $^{1,3}$}
\address{$^1$ School of Computer Science and Technology, Fudan University, Shanghai, China\\
	$^2$  National Institute of Informatics (NII), Tokyo, Japan\\
	$^3$ Shanghai Key Laboratory of Intelligent Information Processing, Fudan University, Shanghai, China}

\begin{document}
	\ninept
	\maketitle
	\begin{abstract}
	Musical audio is generally composed of three physical properties: frequency, time and magnitude. Interestingly, human auditory periphery also provides neural codes for each of these dimensions to perceive music. Inspired by these intrinsic characteristics, a frequency-temporal attention network is proposed to mimic human auditory for singing melody extraction. In particular, the proposed model contains frequency-temporal attention modules and a selective fusion module corresponding to these three physical properties. The frequency attention module is used to select the same activation frequency bands as did in cochlear and the temporal attention module is responsible for analyzing temporal patterns. Finally, the selective fusion module is suggested to recalibrate magnitudes and fuse the raw information for prediction. In addition, we propose to use another branch to simultaneously predict the presence of singing voice melody.  The experimental results show that the proposed model outperforms existing state-of-the-art methods \footnote{Codes are available in https://github.com/yushuai/FTANet-melodic}.
	\end{abstract}
	\section{Introduction}
	\label{sec:intro}
	
	Singing melody extraction estimates the fundamental frequency (F0) of the singing melody, which is one  challenging and critical task in music information retrieval (MIR) \cite{salamon2014melody}. Recently it has become an active research topic with a lot of downstream applications of melody-based AI music, 
	such as cover song identification \cite{serra2010audio}, query-by-humming \cite{WangJ15} and voice separation \cite{IkemiyaYI15}.
	
	With the advances of deep learning technique, several neural network based methods have been proposed to learn the mapping between the audio and melody \cite{kum2016melody, bittner2017deep, basaran2018main, hsieh2018streamlined}. Bittner et al. \cite{bittner2017deep} proposed a fully convolutional neural network to learn the pitch salience and achieved promising results. Hsieh et al. \cite{hsieh2018streamlined} proposed an encoder-decoder architecture and a way to use the bottleneck layer of the network to estimate the presence of a melody line. Unfortunately, other type of network without bottleneck layer cannot enjoy the advantage of it. On the other hand, researchers also attempt to combine the deep learning technique with human auditory \cite{ChouCC18, chen2019cnn, GaoYC20}. For example, Gao et al. \cite{GaoYC20} proposed to integrate a multi-dilation semantic segmentation model with multi-resolution representations to respectively simulate top-down and bottom-up processes of human perception for singing melody extraction. However, prior works treat the three dimensions (i.e., frequency, time and magnitude) equally which is inconsistent with human audition. In addition, directly concatenating the spectral and temporal information may lower the performance of the singing melody extraction. 
	

	Human auditory periphery provides neural codes for frequency, time and magnitude \cite{yost2009pitch}. When the sound enters the cochlea, different frequencies within the sound selectively stimulate different regions of the cochlea.  Then human can extract the pitch via temporal patterns generated by unresolved harmonics in the auditory periphery \cite{Schouten1962}. In this paper, we focus on designing a frequency-temporal attention based model for singing melody extraction to simulate such mechanism. 
	
	Based on the Psycho-acoustic research, the human auditory system will select the stimulated frequency bands in the cochlea. Accordingly, auto-correlation is performed to capture temporal correlations in the auditory cortex. To mimic this mechanism, we introduce frequency attention to assign different weights in the spectrogram along the frequency axis, which corresponds to selecting the stimulated frequency bands in the cochlea. To simulate the process of auto-correlation, a temporal attention module is proposed to capture temporal relations between the adjacent frames, which can capture more complex nonlinear temporal relationship.

	Neuro-physiological research show that some high-frequency signals cannot be perceived by the spectral mode, but temporal models can weakly perceive them \cite{yost2009pitch}. We argue that features from spectral and temporal models cannot be simply concatenated, as the concatenated features may bring noises for melody extraction which hinders the further improvement of this task. Based on extending the idea of selective kernel networks used in computer vision \cite{li2019selective}, we suggest a selective fusion module to dynamically select the features from frequency and temporal attention networks. Accordingly, we fuse the features for thereafter singing melody extraction.

	As shown in Fig.1, in diagram (b), the related frequency bands are enhanced 
	and the unrelated frequency bands are suppressed. Diagram (c) shows the melody line is more salient via the temporal attention. In diagram (d), the selective fusion module dynamically selects and fuses features from (b) and (c) to generate the predicted melody line. We hypothesize that we can learn a frequency-temporal attention based deep architecture for singing melody extraction to simulate these potential intrinsic characteristics. 
	Three technical contributions are made: i) we propose a novel frequency-temporal attention network to mimic the human auditory assigning different weights in the time and frequency axis. To the best of our knowledge, there is no such works for singing melody extraction in the literature. ii) A selective fusion module is proposed to dynamically assign weight to the spectral features and temporal features.  Then we fuse them for melody extraction. iii) We propose to use another branch to directly predict the presence of melody. 
	
	
	\begin{figure}[ht]
		\centering
		\begin{subfigure}[b]{0.23 \textwidth}
			\includegraphics[height=2cm,width=4cm]{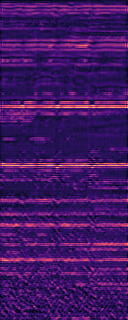}
			\caption{Input Spectrogram.}
		\end{subfigure}
		\begin{subfigure}[b]{0.23 \textwidth}
			\includegraphics[height=2cm,width=4cm]{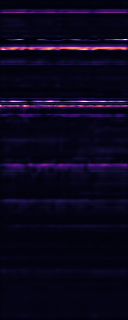}
			\caption{Frequency Attention Output.}
		\end{subfigure}
		\begin{subfigure}[b]{0.23 \textwidth}
			\includegraphics[height=2cm,width=4cm]{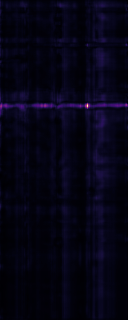}
			\caption{Temporal Attention Output.}
		\end{subfigure}
		\begin{subfigure}[b]{0.23 \textwidth}
			\includegraphics[height=2cm,width=4cm]{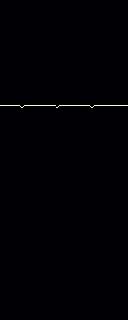}
			\caption{Selective Fusion Output.}
		\end{subfigure}
		
		\caption{Procedures of singing melody extraction of the 128-256 ms of `daisy4.wav' in the ADC2004 dataset performed by the proposed model.}
		\label{fig:proc}
	\end{figure}
	
	\section{Proposed model}
	The overall architecture of our frequency-temporal attention network is shown
	in Fig. 2. It has two branches:  the top branch contains stacked convolution layers performing
	fast downsampling to obtain high-level semantic representation for singing melody detection; and the bottom branch contains the proposed frequency-temporal attention and selective fusion modules. The  Frequency attention and temporal attention, selective fusion module, singing melody detection in the proposed deep architecture are respectively addressed in the Section. 
	\subsection{Model Input}
	We choose Combined Frequency and Periodicity (CFP) representation \cite{su2015combining} as the input of our model due to its effectiveness and popularity. The CFP representation has three parts: the power-scaled spectrogram, generalized cepstrum (GC) \cite{kobayashi1984spectral, tokuda1994mel}  and generalized cepstrum of spectrum (GCoS) \cite{su2017between}. We use 44100 Hz sampling rate, 2,048-sample window size, and 256-sample hop size for computing the short-time-Fourier-transformation (STFT).
	
	\subsection{Frequency-temporal Attention Module}
	The design of frequency-temporal attention module represents the first contribution of this work.  The detailed architecture is shown in Fig. 2. Unlike previous works for speech-related tasks, we do not employ classic attention models \cite{bahdanau2014neural,vaswani2017attention}. Instead we use 1-D convolution to select relevant regions (frequency/time) for this task. 
	
	Formally, given the input feature map $\mathbf{S}$  $\in \mathbb{R}^{F\times T \times C'}$, an average pooling is firstly applied to the spectrogram for calculating the distribution of magnitudes along the time axis. Unlike 2-D average pooling, we use row average pooling to achieve this. The frequency descriptor $f \in \mathbb{R}^{F \times C'}$ can be calculated:
	\begin{equation}
	f = \frac{1}{T}\sum_{i<=T} s_{ij}
	\end{equation}
	where $s_{ij}$ is the element in the $i$-th row and $j$-th column in $\mathbf{S}$. 
	
	\begin{figure}[ht]
		\centering
		\includegraphics[width=8.5cm]{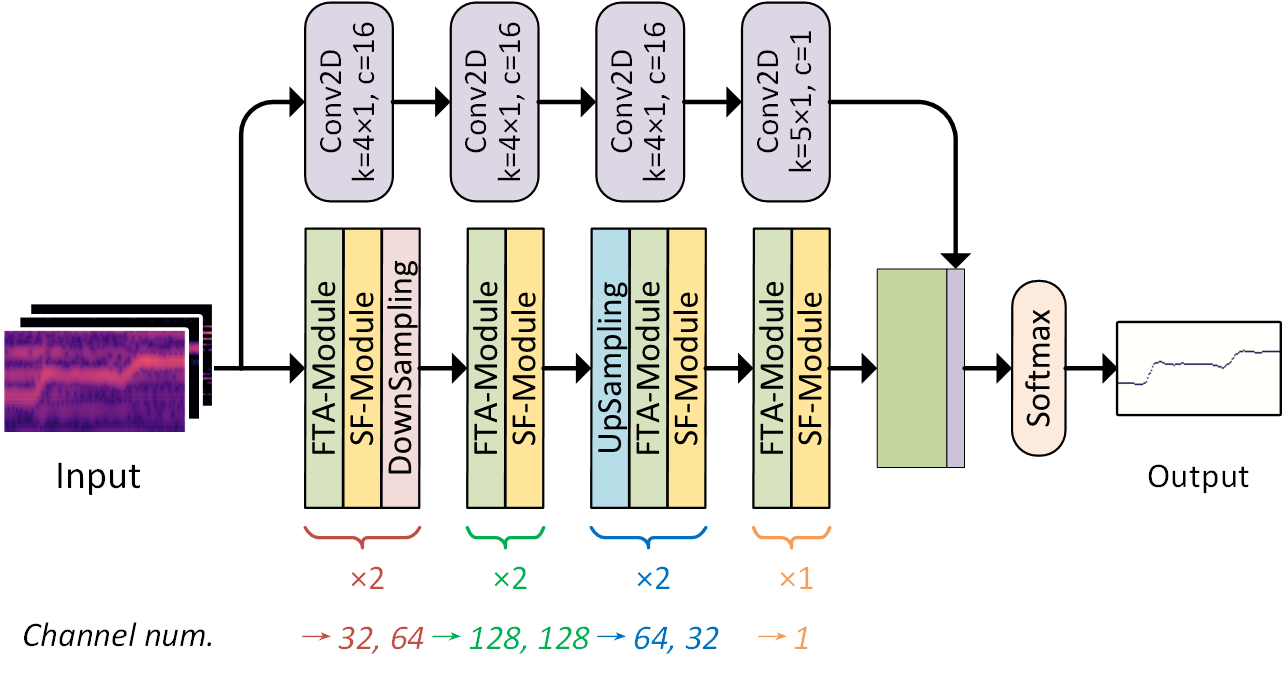}
		\caption{The overall architecture of the proposed model. The top branch is proposed to predict the presence of a melody. The bottom branch consists of the proposed frequency-temporal attention module and selective fusion module for F0 estimation. `k' and `c' denotes for the kernel size and the number of channels, respectively. }
		\label{fig:arch}
	\end{figure}
	\begin{figure}
		\centering
		\includegraphics[width=7.5cm]{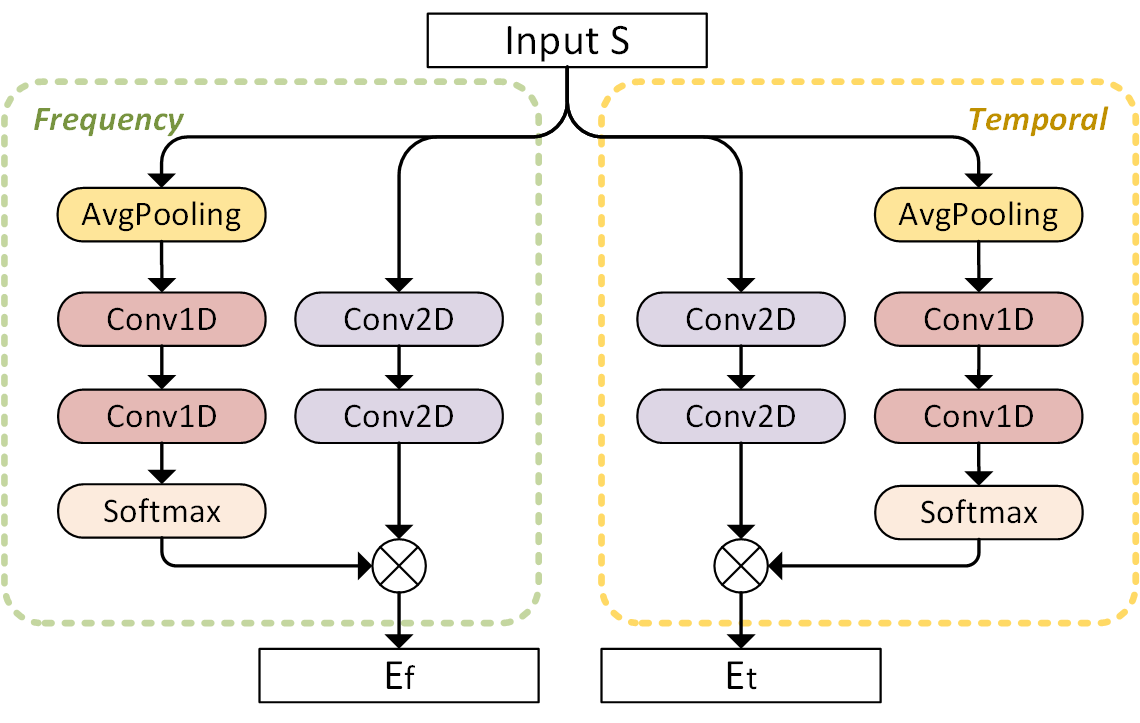}
		\caption{The detailed architecture of the proposed frequency-temporal attention module. $
		E_f$ and $E_t$ are the output of frequency attention and temporal attention, respectively. }
		\label{fig:tfa}
	\end{figure}
	
	Then 1-D convolution is used to dynamically select task-related frequency for singing melody extraction. Since 1-D convolution is good at learning the relationship along the frequency bins or time axis, we choose to employ 1-D convolution to perform frequency-temporal attention. For a frequency descriptor $f$, the process of 1-D convolution can be written as:
	\begin{equation}
	V^l=f * k^l
	\end{equation}
	where $k^l$ is the kernel of 1-D convolution at $l$-th layer and $V^l \in \mathbb{R}^{F \times C}$ is the newly generated feature map, * stands for the convolution operation. Finally, we apply a softmax layer to the output of the 1-D convolution layer $V$ to obtain the frequency attention map $\mathbf{A_f}\in \mathbb{R}^{F \times C}$:
	\begin{equation}
	\mathbf{A_f} = softmax(V)
	\end{equation}
	Similarly, we can also obtain the temporal attention map $\mathbf{A_t}\in \mathbb{R}^{T \times C}$ through the same process mentioned above. 
	
	Meanwhile, we feed feature map $\mathbf{S}$ into 2-D convolution layers with the kernel size of $(3 \times 3)$ and $(5 \times 5)$ to generate two new feature maps $\{\mathbf{S_f}, \mathbf{S_t}\}$$ \in \mathbb{R}^{F\times T \times C}$. Then we perform matrix multiplications between $\{\mathbf{S_f}, \mathbf{S_t}\}$ and $\{\mathbf{A_f}, \mathbf{A_t}\}$ to obtain the final output $\mathbf{E}=\{\mathbf{E_f}, \mathbf{E_t}\}$:
	\begin{equation}
	\begin{split}
	\mathbf{E_f} &= \mathbf{S_f} \otimes broadcast(\mathbf{A_f}), \\
	\mathbf{E_t} &= \mathbf{S_t} \otimes broadcast(\mathbf{A_t}).
	\end{split}
	\end{equation}
	where $broadcast$ is the process of making matrices with different shapes have compatible shapes for element-wise multiplication.
	\begin{figure}
		\centering
		\includegraphics[width=7.9cm, height=4.5cm]{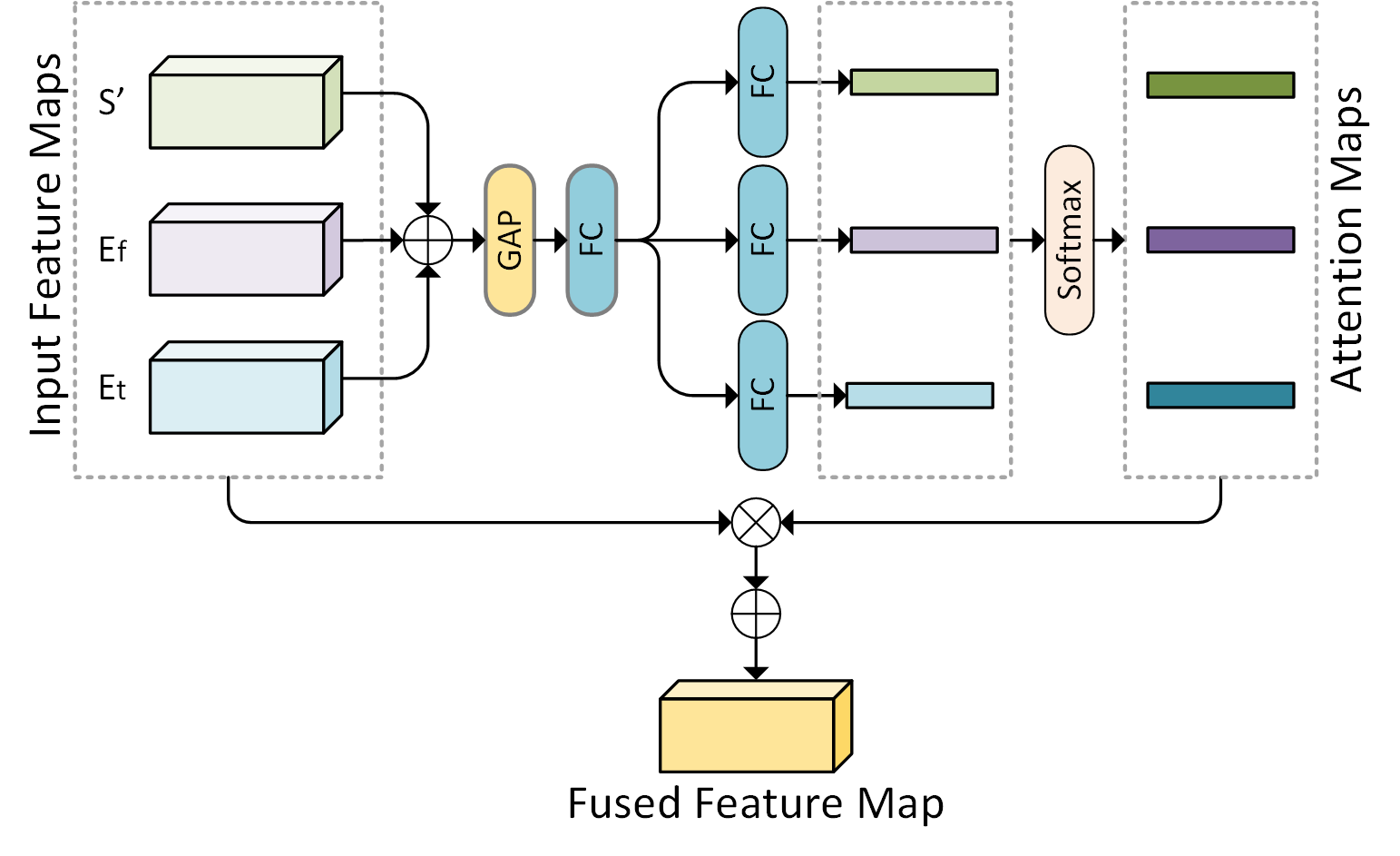}
		\caption{The detailed architecture of the proposed selective fusion module.}
		\label{fig:sfm}
	\end{figure}
	\subsection{Selective Fusion Module}
	The design of selective fusion module (SFM) represents the second contribution of this work. Inspired by the idea of selective kernel networks used in computer vision \cite{li2019selective}, we devise this module to dynamically select spectral and temporal features and fuse them. The detailed architecture is shown in Fig. 3. This module takes three inputs: the feature map $\mathbf{S'} \in \mathbb{R}^{F \times T \times C}$  generated from $\mathbf{S}$ via a $(1\times 1)$ convolution, the frequency attention result $\mathbf{E_f}$ and the temporal attention result $\mathbf{E_t}$. Firstly, an element-wise addition is performed to fuse the three inputs into a new feature map $\mathbf{\Gamma}$ and then a global average pooling (GAP) is performed to obtain global descriptor $g \in \mathbb{R}^{C}$. 
	\begin{equation}
	g=\frac{1}{F\times T} \sum_{i<=F,  j<=T}\Gamma_{ij}
	\end{equation}
	After a fully connected (FC) layer for non-linear transformation, three FC layers are used to learn the importance of each channel of the feature maps. The softmax layer is applied to obtain the attention map. After obtaining the attention maps, matrix multiplication is performed between the three inputs and the attention maps to obtain the weighted feature maps. Finally, we fused the three weighted feature maps by element-wise addition operation. The fuse feature map contains rich information selected from the spectral and temporal modes.
	
	\subsection{Melody Detection Branch}   
	The design of melody detection branch (MDB) represents the third contribution of this work. The motivation of this branch is to design a general scheme that does not depend on special structure and can fast predict the presence of a melody. In this branch, we employ four stacked convolution layers to directly perform downsampling on the spectrograms. Specifically, we have the first three convolution layers with the kernel size $(4 \times 1)$ and stride size 
	$(4 \times 1)$ and the last convolution layer with the kernel size $(5 \times 1)$  and stride size 
	$(5 \times 1)$. As a result, the output of this branch is $m \in \mathbb{R}^{1\times T}$ stride size of $(5 \times 1)$. Following \cite{hsieh2018streamlined}, the output of MDB is concatenated with the salience map to make an $(F+1) \times T$ matrix. By including this branch, the voicing recall (VR) rate can be improved.  
	
	\section{Experiments}
	\begin{table}[h]
		\captionsetup[subfloat]{labelfont=bf}
		\centering
		\begin{subtable}[t]{0.79\linewidth}
			\begin{tabular}{|c|l|l|l|l|l|}
				\hline
				\centering
				\multirow{2}{*}{Method} & \multicolumn{5}{c|}{ADC2004 (vocal)} \\ \cline{2-6}
				& OA & RPA & RCA & VR & VFA  \\ \hline
				
				w/o F-att.  & 83.9 & 82.4 & 84.2 & 84.1 & 8.8 \\ \hline
				w/o T-att. & 83.5 & 82.3  & 84.7 & 85.2 & 8.4 \\ \hline
				w/o SFM & 84.5 & 83.4 & 85.1 & 85.2 & \textbf{8.2} \\ \hline
				w/o MDB & 84.7 & 84.0 & 85.6 & 87.1 & 11.3 \\ \hline
				\textbf{Proposed} & \textbf{85.9} & \textbf{85.2} & \textbf{86.7} & \textbf{87.5} & 11.1 \\ \hline
			\end{tabular}
			\caption{\textbf{ADC2004}}
		\end{subtable}
		
		\begin{subtable}[t]{1\linewidth}
			\centering
			\begin{tabular}{|c|l|l|l|l|l|}
				\hline
				\multirow{2}{*}{Method} & \multicolumn{5}{c|}{MIREX 05 (vocal)}  \\ \cline{2-6}
				& OA & RPA & RCA & VR & VFA  \\ \hline
				w/o F-att. & 81.7 & 76.5 & 77.7 & 81.0 & \textbf{9.2} \\ \hline
				w/o T-att. &  78.3 & 78.0 & 79.0 & 84.0 & 20.2\\ \hline
				w/o SFM & 79.3 & 74.9 & 76.0 & 78.8 & 13.2 \\ \hline
				w/o MDB & 81.0 & 79.6 & 80.6 & 84.6 & 16.6 \\ \hline
				\textbf{Proposed} & \textbf{84.0} & \textbf{80.0} & \textbf{80.8} & \textbf{85.2} & 9.7\\ \hline
			\end{tabular}
			\subcaption{\textbf{MIREX 05}}
		\end{subtable}
		
		\caption{Results of Ablation Study on ADC2004 and MIREX 05 dataset. The values in the table are percentile. ``w/o F-att'' and ``w/o T-att'' denote for without frequency/temporal attention respectively. ``w/o SFM'' stands for without selective fusion module. ``w/o MDB'' stands for without melody detection branch.}
		\label{rq21}
	\end{table} 
	\begin{table}[h]
		\captionsetup[subfloat]{labelfont=bf}
		\centering
		\begin{subtable}[t]{0.82\linewidth}
			\begin{tabular}{|c|l|l|l|l|l|}
				\hline
				\centering
				\multirow{2}{*}{Method} & \multicolumn{5}{c|}{ADC 2004 (vocal)} \\ \cline{2-6}
				& OA & RPA & RCA & VR & VFA  \\ \hline
				MCDNN \cite{kum2016melody}& 69.8 & 70.2 & 74.3 & 79.0 & 30.7 \\ \hline
				SegNet \cite{hsieh2018streamlined} & 81.6 & 82.7 & 84.9 & 87.4 & 18.6 \\ \hline
				MD+MR \cite{GaoYC20} & 82.8 & 82.4 & 84.6 & 85.3 & 12.7 \\ \hline
				\textbf{Proposed} & \textbf{85.9} & \textbf{85.2} & \textbf{86.7} & \textbf{87.5} & \textbf{11.1} \\ \hline
			\end{tabular}
			\caption{\textbf{ADC 2004}}
		\end{subtable}
		\hfill
		\begin{subtable}[t]{0.82\linewidth}
			\centering
			\begin{tabular}{|c|l|l|l|l|l|}
				\hline
				\multirow{2}{*}{Method} & \multicolumn{5}{c|}{MIREX 05 (vocal)}  \\ \cline{2-6}
				& OA & RPA & RCA & VR & VFA  \\ \hline
				MCDNN \cite{kum2016melody} &  75.6 & 69.4 & 71.1 & 74.9 & 14.7 \\ \hline
				SegNet \cite{chen2019cnn} & 78.6 & 78.4 & 79.7 & 85.7 & 21.5 \\ \hline
				
				MD+MR \cite{GaoYC20} & 80.7 & 78.6 & 79.5 & 84.3 & 16.9 \\ \hline
				\textbf{Proposed} & \textbf{84.0} & \textbf{80.0} & \textbf{80.8} & \textbf{85.2} & \textbf{9.7}\\ \hline
			\end{tabular}
			\caption{\textbf{MIREX 05}}
		\end{subtable}
		\hfill
		\begin{subtable}[t]{0.82\linewidth}
			\centering
			\begin{tabular}{|c|l|l|l|l|l|}
				\hline
				\multirow{2}{*}{Method} & \multicolumn{5}{c|}{Medley DB (vocal)}  \\ \cline{2-6}
				& OA & RPA & RCA & VR & VFA  \\ \hline
				MCDNN \cite{kum2016melody} &  60.0 & 52.2 & 56.3 & 59.5 & 19.6\\ \hline
				SegNet \cite{hsieh2018streamlined} & 65.8 & \textbf{59.3} & \textbf{64.8} & \textbf{72.1} & 27.9 \\ \hline
				
				MD+MR \cite{GaoYC20} & 62.3 & 52.2 & 57.4 & 62.9 & \textbf{21.5} \\ \hline
				\textbf{Proposed} & \textbf{66.3} & 58.0 & 63.5 & 69.3 & 23.6\\ \hline
			\end{tabular}
			\subcaption{\textbf{MedleyDB}}
		\end{subtable}
		
		\caption{Results of the proposed and baseline methods on ADC 2004, MIREX 05 and Medley DB dataset. The values in the table are percentile.}
		\label{tab2}
	\end{table} 
	\subsection{Experiment Setup}
	We randomly choose 60 vocal tracks from the Medley DB dataset. To increase the amount of the training data, we augment the training dataset by copying some of the chosen vocal tracks. Accordingly, there are 98 clips in the training dataset.  We select only samples having melody sung by human voice from ADC2004, MIREX 05\footnote{https://labrosa.ee.columbia.edu/projects/melody} and MedleyDB for test sets. As a result, 12 clips in ADC2004, 9 clips in MIREX 05 and 12 clips in MedleyDB  are selected. Note that there is no overlap between the training and testing datasets.
	
	To adapt the pitch ranges required in singing melody extraction, following \cite{hsieh2018streamlined} we set hyperparameters in computing the CFP for our model. For vocal melody extraction, the number of frequency bins is set to 320, with 60
	bins per octave, and the frequency range is from 31 Hz ($B0$) to
	1250 Hz (${\settoheight{\dimen0}{D}D\kern-.05em \resizebox{!}{\dimen0}{\raisebox{\depth}{$\sharp$}}}6$). We divide the training clips into fixed-length segments of T = 128 frames, which is 128 milliseconds.  Our model is implemented with Keras \footnote{https://keras.io}. For model update, we choose the binary cross entropy as the loss function. The Adam optimizer is used with the learning rate of 0.0001.
	
	Following the convention in the literature, we use the following metrics for performance evaluation: overall accuracy (OA), raw pitch accuracy (RPA), raw chroma accuracy (RCA), voicing recall (VR) and voicing false alarm (VFA). We use mir\_eval library \cite{raffel2014mir_eval} with the default setting to calculate the metrics. For each metric other than VFA, the higher score, the higher performance. In the literature \cite{salamon2014melody}, OA is often considered more important than other metrics.
	\subsection{Ablation Study}
	To investigate how much the proposed frequency-temporal attention module contributes to the model, we first remove the frequency attention, and only temporal attention is used to encode the temporal information. As is shown in Table 1, the performances on both datasets are decreased. When focusing on OA, the performance is decreased by 2.4\% on ADC2004 and by 2.8\% on MIREX 05. We then remove the temporal attention and only keep the frequency attention. The performances on both datasets are decreased by 2.9\% and by 7.3\%, respectively, on the both datasets.
	
	We then investigate the effectiveness of the selective fusion module. When focusing on OA, the performances of the ablated version are decreased by 2.8\% and 5.9\% on ADC2004 and MIREX 05, respectively. The results justify the assumption that direct concatenation may hinder the further improvement of the model. Lastly, we investigate the effectiveness of the proposed melody detection branch. The results w.r.t. OA on both datasets are decreased by 1.4\% and 3.7\%, respectively, on the two datasets. The results demonstrate the effectiveness of the proposed simple design for melody detection. However, when focusing on VFA, the ablated version achieves a better score than the proposed model. We analyzed this phenomenon, and found that the shallow CNNs cannot capture semantic information well and it is prone to predict a non-melody frame with complex harmonic patterns as a melody one. 
	\begin{figure}[t]
		\centering
		\begin{subfigure}[b]{0.23 \textwidth}
			\includegraphics[scale=0.27]{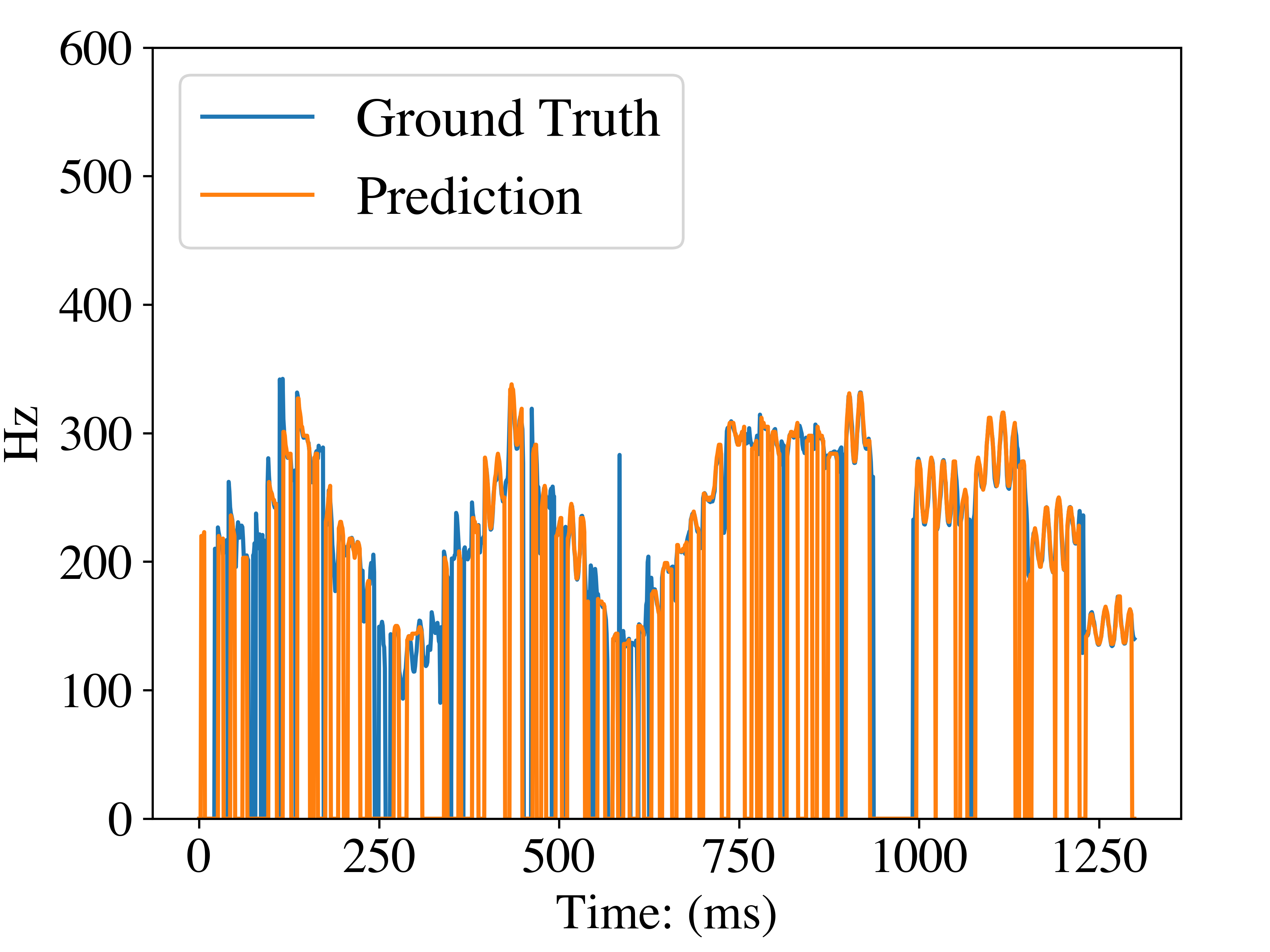}
			\caption{Opera\_male3 with ours.}
		\end{subfigure}
		\begin{subfigure}[b]{0.23 \textwidth}
			\includegraphics[scale=0.27]{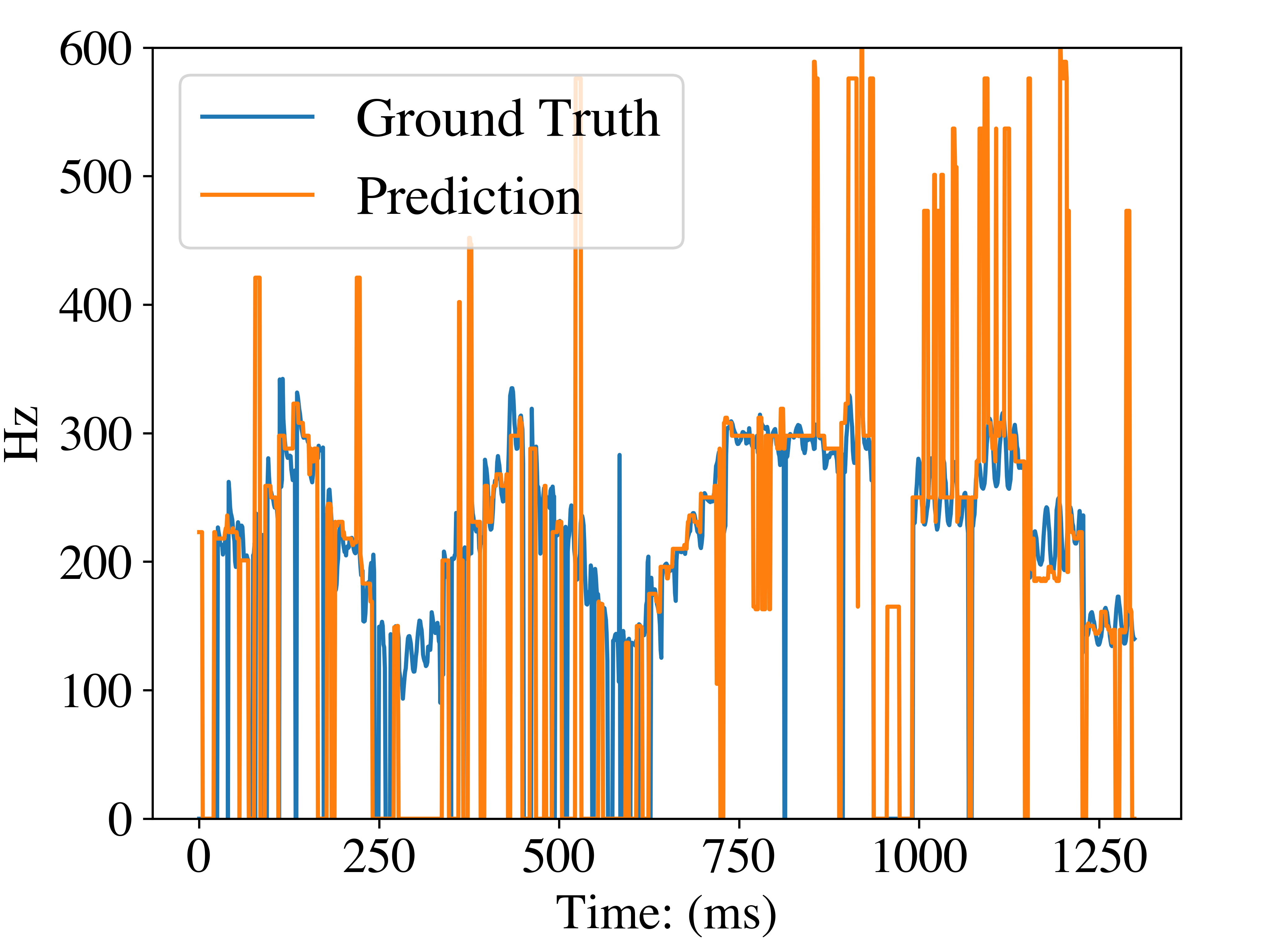}
			\caption{Opera\_male3 with SegNet.}
		\end{subfigure}
		\begin{subfigure}[b]{0.23 \textwidth}
			\includegraphics[scale=0.27]{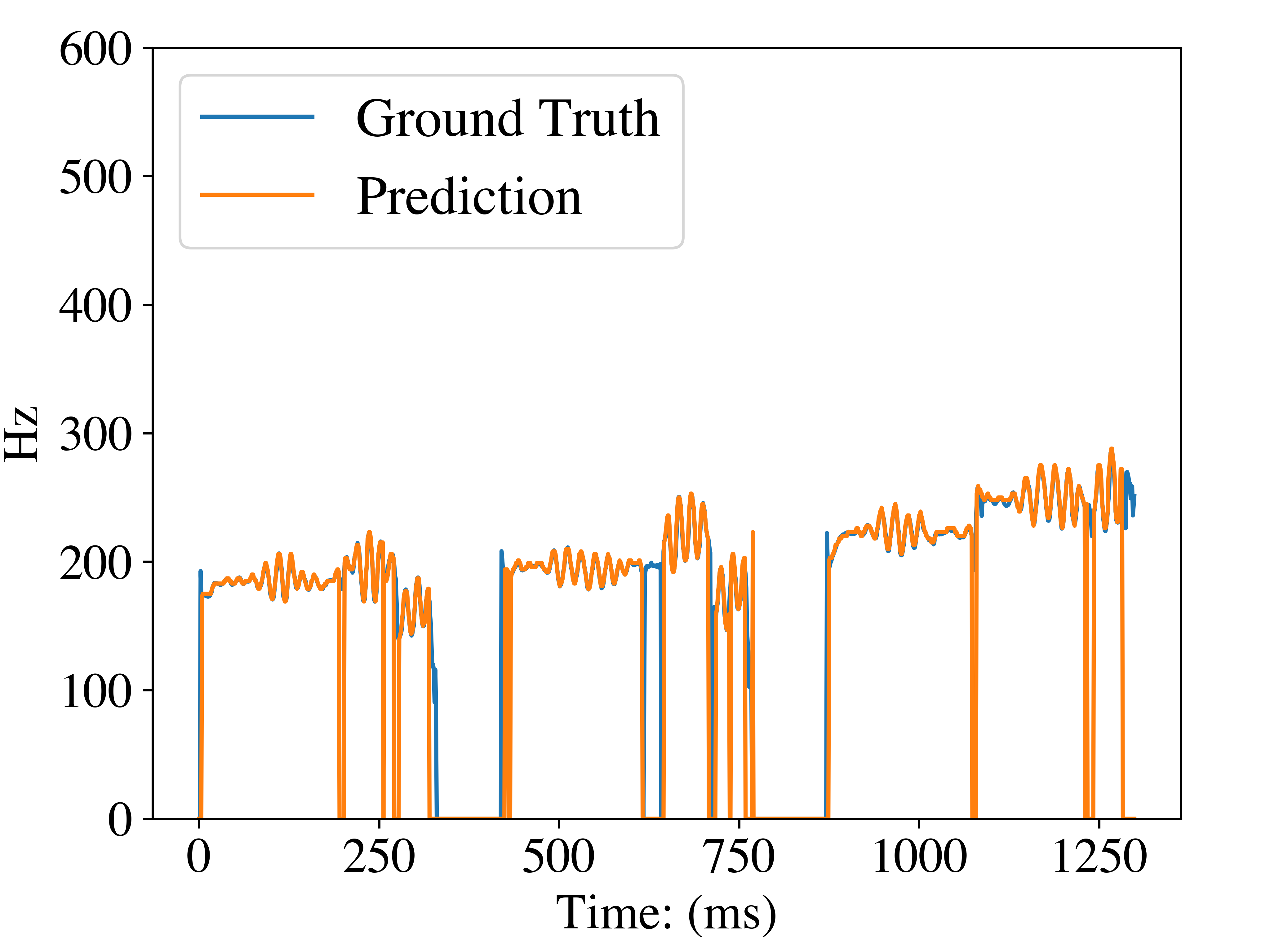}
			\caption{Opera\_male5 with ours.}
		\end{subfigure}
		\begin{subfigure}[b]{0.23 \textwidth}
			\includegraphics[scale=0.27]{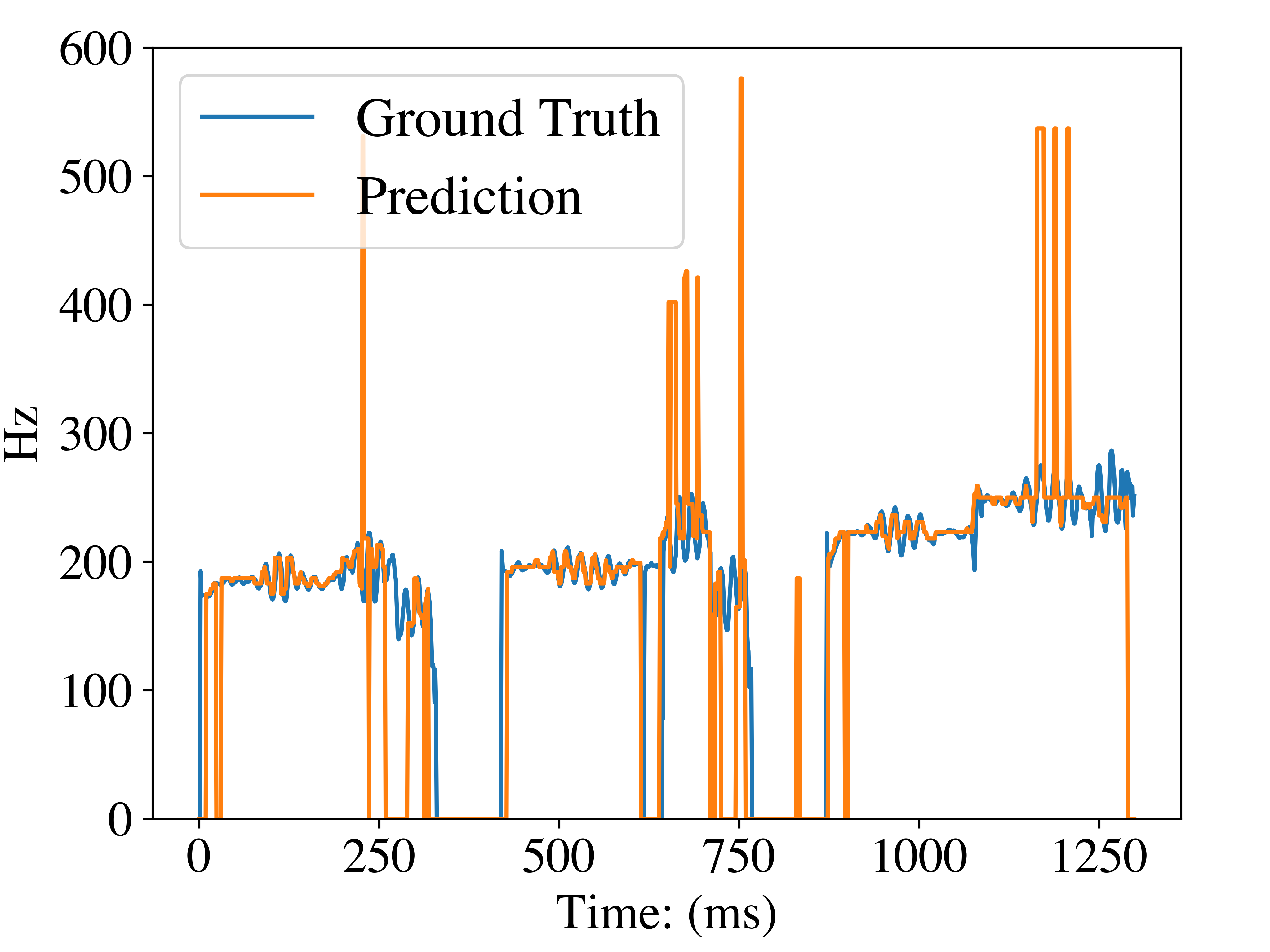}
			\caption{Opera\_male5 with SegNet.}
		\end{subfigure}
		
		\caption{Visualization of singing melody extraction results on two opera songs using different models.}
		\label{visual_plot}
	\end{figure}
	\subsection{Comparison with existing works}
	The performances on the three datasets are listed in Table 2. Three baseline methods are compared in Table 2, including MCDNN \cite{kum2016melody}, SegNet \cite{hsieh2018streamlined} and MD+MR \cite{GaoYC20}. We carefully tune the hyper-parameters of the three baseline methods to ensure that they reach their peak performances on our training dataset. The proposed model and the three baseline methods are trained on the same dataset. Compared with the baseline methods, the proposed method achieves the highest score in general. The results clearly confirm the effectiveness and robustness of our proposed model. For comparison with other baselines, when focusing on OA, the proposed method outperforms MCDNN by 23.1\% in ADC2004, by 11.1\% in MIREX 05, and by 10.5\% in Medley DB. Since all of the three baselines have melody-detectors, when focusing on VR and VFA, the proposed method achieves the highest score in ADC2004 and MIREX 05 datasets, and comparable result in MedleyDB. 
	
	To investigate what types of errors are solved by the proposed model, a case study is performed on two opera songs: ``opera\_male3.wav'' and ``opera\_male5.wav'' in the ADC2004 dataset. We choose SegNet \cite{hsieh2018streamlined} to compare with due to its effectiveness and popularity. As depicted in Fig. 5, we can observe that there are fewer octave errors in diagram (a) and (c) than in diagram (b) and (d). Moreover, from 1000-1200 ms in the diagram (d), we can find some errors that predict the wrong frequency bin near the right one, which are correctly predicted in diagram (c). Through the visualization of the predicted melody contour, we can say that the performance gains of the proposed model can be attributed to solving the octave errors and other errors. However, we can also observe that there seem to be more the melody detection errors (i.e., predicting a non-melody frame as a melody one) than in SegNet \cite{hsieh2018streamlined}. 
	The goal of the melody detection branch is to fast predict the presence of the melody which does not depend on special structure. We leave this for future research topic to design more accurate and fast network for improving the quality of melody detection.

	\section{Conclusion}
	In this paper, we propose a novel frequency-temporal attention network to mimic the human auditory for singing melody extraction, which mainly contains three novel modules: frequency-temporal attention, selective fusion, and singing melody detection. Frequency-temporal feature learning and singing melody detection are simultaneously learned in an end-to-end way. Experimental results show the proposed model outperforms the existing state-of-the-art models on three datasets. Designing more accurate and faster method to improve the performance of singing melody detection will be our future work.
	\section{Acknowledgement}
This work was supported by National Key R\&D Program of China (2019YFC1711800), NSFC (61671156).
	
	\bibliographystyle{IEEEbib}
	\bibliography{strings,refs}
	
\end{document}